\documentclass[12pt]{article}

\begin{document}
\textbf{VISCOUS  COSMOLOGY  AND  THE  CARDY-VERLINDE  FORMULA}\\
\bigskip

\begin{center}
Iver Brevik\footnote{Email: iver.h.brevik@mtf.ntnu.no. Contribution to the Proceedings of the Second Londrina Winter School "Mathematical Methods in Physics", August 25-30, 2002, Londrina-Parana, Brazil.}

\bigskip

Department of Energy and Process Engineering, Norwegian University of Science and Technology, N-7491 Trondheim, Norway\\

\bigskip

January 2003
\end{center}

\begin{abstract}
The holographic principle in a radiation dominated universe is extended to incorporate the case of a bulk-viscous cosmic fluid. This corresponds to a nonconformally invariant theory. Generalization of the Cardy-Verlinde entropy formula to the viscous case appears to be possible from a formal point of view, although we question on physical grounds the manner in which the Casimir energy is evaluated in this case. Also, we consider an observation recently made by Youm, namely that the entropy of the universe is no longer expressible in the conventional Cardy-Verlinde form if one relaxes the radiation dominance equation of state and instead merely assumes that the pressure is proportional to the energy  density. We show that Youm's generalized entropy formula remains valid when the cosmic fluid is no longer ideal, but endowed with a constant bulk viscosity.
\end{abstract}

\section{Introduction}	

The suggestion of Verlinde \cite{verlinde00} that there exists a holographic bound on the subextensive entropy associated with the Casimir energy makes it natural to ask: Is this formal merging between the holographic principle, the Cardy entropy formula from conformal field theory (CFT) \cite{cardy86,blote86}, and the Friedmann equations from cosmology, only a formal coincidence or does it reflect a deep physical property? As one would expect, the Verlinde suggestion has given rise to a large number of investigations, where various facets of the Cardy-Verlinde entropy formula, in particular, have been discussed. An extensive list of recent references can be found, for instance, in the paper of Youm \cite{youm02}. Let us also mention the large interest that has arisen in connection with entropy and energy as following from quantum and thermal fluctuations in conformal field theories \cite{kutasov01}.

Now Verlinde assumed a spatially closed, radiation dominated universe. It becomes natural to inquire to what degree these basic assumptions can be generalized. The present paper focuses on one specific generalization, namely the presence of a {\it bulk viscosity} in the early universe. This corresponds to an extension of the formalism to nonconformally invariant theories. Viscous cosmology theories as such have recently attracted some interest - cf., for instance, Ref.~\cite{harko02}.

We will consider two topics:

(i)  Generalization of the Cardy-Verlinde formula to the case of a constant bulk viscosity in the cosmic fluid \cite{brevik02}.

(ii)  Generalization of Youm's entropy formula \cite{youm02,brevik02a} to the constant bulk-viscosity case. We here assume, as did Youm, that the conventional radiation-dominance equation of state is replaced with a more general equation 
\begin{equation}
p=(\gamma -1)\rho ,
\end{equation}
\label{1}
with $\gamma$ a constant in the interval $0\le \gamma \le 2$.

The cosmological constant $\Lambda$ will for the most part be taken to be zero, although we briefly comment on the case of $\Lambda \neq 0$ in section 2.3.

We use natural units, with $\hbar =c=k_B=1$.

\section{Cardy-Verlinde formula for the viscous fluid}

\subsection{Relationship to the Cardy-Verlinde formula in CFT}

Consider the cosmic fluid whose four-velocity is 
$U^\mu=(U^0, U^i)$. In comoving coordinates, $U^0=1,~U^i=0$. In terms of the projection tensor $h_{\mu \nu}=g_{\mu \nu}+U_\mu U_\nu$ we can write the fluid's energy-momentum tensor as
\begin{equation}
T_{\mu\nu}=\rho U_\mu U_\nu+(p-\zeta \theta)h_{\mu\nu}-2\eta \sigma_{\mu\nu},
\end{equation}
\label{2}
assuming constant temperature in the fluid. Here, $\zeta$ is the bulk viscosity, $\eta$ the shear viscosity, $\theta \equiv {U^\mu}_{; \mu}$ the scalar expansion, and $\sigma_{\mu\nu}=h_\mu ^\alpha \,h_\nu ^\beta\, U_{(\alpha; \beta)}-\frac{1}{3}h_{\mu\nu} \,\theta$ the shear tensor. In accordance with common usage we omit henceforth the shear viscosity in view of the assumed complete isotropy of the fluid, although we have to mention that this is actually a nontrivial point. The reason is that the shear viscosity is usually so much greater than the bulk viscosity. (Typically, after termination of the plasma era at the time of recombination ($T \simeq 4000$ K) the ratio $\eta / \zeta$ as calculated from kinetic theory is as large as about $10^{12}$  \cite{brevik94}. Thus, even a slight anisotropy in the fluid would easily outweigh the effect of the minute bulk viscosity.)

Assuming now a metric of the FRW type,
\begin{equation}
ds^2=-ds^2+a^2(t)\left( \frac{dr^2}{1-kr^2}+r^2 d\Omega^2 \right),
\end{equation}
\label{3}
with $k=-1,0,1$ the curvature parameter, we get $\theta =3\dot{a}/a \equiv 3H$. The effective pressure $\tilde{p}$ is
$\tilde{p} \equiv p-\zeta \theta =p-3H\zeta.$
From Einstein's equations $R_{\mu\nu}-\frac{1}{2}Rg_{\mu\nu}+\Lambda g_{\mu\nu}=8\pi G\,T_{\mu\nu}$ we obtain the first Friedmann equation ("initial value equation")
\begin{equation}
H^2=\frac{8\pi G}{3}\,\rho+\frac{\Lambda}{3}-\frac{k}{a^2},
\end{equation}
\label{4}
where $\rho=E/V$ is the  energy density. This equation contains no viscous term. The second Friedmann equation ("dynamic equation"), when combined with Eq.~(4), yields
\begin{equation}
\dot{H}=-4\pi G (\rho +\tilde{p})+\frac{k}{a^2},
\end{equation}
\label{5}
in which the presence of viscosity is explicit.

We recall that the entropy of a (1+1) dimensional CFT is given by the Cardy formula \cite{cardy86,blote86}
\begin{equation}
S=2\pi \sqrt{(c/6)(L_0-c/24)},
\end{equation}
\label{6}
where $c$ is the central charge and $L_0$ the lowest Virasoro generator.

  Let us assume that the universe is closed, and has a vanishing cosmological constant, $k=+1,~\Lambda=0$.
This is the case considered in  \cite{verlinde00} (in his formalism the number $n$ of space dimensions is set equal to 3).  The Friedmann equation (4) is seen to agree with the CFT equation (6) if we perform the substitutions
\begin{equation}
 L_0 \rightarrow Ea/3,~~~c\rightarrow 3V/(\pi Ga),~~~S\rightarrow HV/(2G).
\end{equation}
\label{7}
These substitutions are the same as in Ref.\cite{verlinde00}. We can thus conclude: {\it The formal comparison made by Verlinde with the CFT formula goes through unchanged even if the fluid possesses a bulk viscosity}. Note that no assumptions have so far been made about the equation of state for the fluid.

\subsection{Further entropy considerations}

Let us have a closer look at the entropy concept in the presence of viscosity. In the conventional nonviscous theory there are actually three different entropy definitions.  First, there is the Bekenstein entropy \cite{bekenstein81},  $S_B=(2\pi/3)Ea$. The arguments for deriving this expression seem to be of a general nature; in accordance with Verlinde we find it likely that the Bekenstein bound $S \leq S_B$ is universal. We shall accept this expression for $S_B$ in the following, even when the fluid is viscous.

The next kind of entropy is the Bekenstein-Hawking expression $S_{BH}$, which is supposed to hold for systems with limited self-gravity: $S_{BH}=V/2Ga.$ Again, this expression relies upon the viscous-insensitive member (4) of Friedmann's equations. Namely, when $\Lambda=0$ this equation yields
 $ S_B  < S_{BH}$ when $ Ha < 1$ and $S_B > S_{BH}$ when $Ha >1$. 
The borderline case between a weakly and a strongly gravitating system is 
thus at $Ha=1$. It is reasonable to identify $S_{BH}$ with the holographic 
entropy of a black hole with the size of the universe.

The third entropy concept is the Hubble entropy $S_H$. It can be introduced by starting from the conventional formula $A/4G$ for the entropy of a black hole. The horizon area $A$ is approximately $H^{-2}$, so that $S_H \sim H^{-2}/4G \sim HV/4G$ since $V\sim H^{-3}$. Arguments have been given by several researchers \cite{easther99}
 for assuming the maximum entropy inside the universe to be produced by black holes of the size of the Hubble radius. According to Verlinde the FSB prescription (see \cite{verlinde00} for a closer discussion) one can determine the prefactor:  $S_H=HV/2G$. It is seen to agree with Eq.~(7).

One may now {\it choose} (see item (ii) below) to define the Casimir energy $E_C$ as the violation of the Euler identity:
\begin{equation}
E_C \equiv 3(E+pV-TS)
\end{equation}
\label{8}
where, from scaling, the total energy $E$ can be decomposed as ($E_E$ is the extensive part) $E(S,V)=E_E(S,V)+\frac{1}{2}E_C(S,V)$. Due to  conformal invariance the products $E_E\, a$
and $E_C\, a$ are independent of the volume $V$, and a function of the
entropy $S$ only. From the known extensive behaviour of $E_E$ and the
sub-extensive behaviour of $E_C$ one may write (for CFT)
 \begin{equation}
E_E= (C_1/4\pi a) S^{4/3},~~~~E_C= (C_2/2\pi a) S^{2/3},
\end{equation}
\label{9}
where $C_1, C_2$ are constants whose product for CFTs is known: $\sqrt{C_1
C_2}=n=3$ (this follows from the AdS/CFT correspondence, cf. \cite{verlinde00}). From these expressions it follows that
\begin{equation}
S=(2\pi/3) a\sqrt{E_C(2E-E_C)}.
\end{equation}
\label{10}
This is the Cardy-Verlinde formula. Identifying $Ea$ with $L_0$ and $E_C\,a$ with $c/12$ we see that Eq.~(10) becomes the same as Eq.~(6), except from a numerical prefactor which is related to our assumption about $n=3$ space dimensions instead of $n=1$ as assumed in the Cardy formula.

The question is now: can the above line of arguments be carried over to
the case of a viscous fluid? The most delicate point here appears to be
the assumed pure entropy dependence of the product $Ea$. As we mentioned
above, this property was derived from conformal invariance, a property that is 
absent in the case under discussion. To examine
whether the property still holds when the fluid is viscous (and conformal
invariance is lost), we can start from the Friedmann equations (4) and (5), in the case $k=1, \, \Lambda=0$, and derive the "energy equation", which can be transformed to
\begin{equation}
\frac{d}{da}(\rho a^4)=(\rho -3\tilde{p})a^3.
\end{equation}
\label{11}
Thus, for a radiation dominated universe, $p=\rho /3$, it follows that
\begin{equation}
\frac{d}{dt}(\rho a^4)=\zeta \,\theta^2 a^4.
\end{equation}
\label{12}
Let us compare this expression, which is essentially the time derivative of the volume density of the quantity $Ea$ under discussion, with the four-divergence of the entropy current four-vector $S^\mu$. If $n$ is the number density and $\sigma$ the entropy per particle, we have $S^\mu=n\sigma U^\mu$, which satisfies the relation (cf., for instance, Ref.~\cite{brevik94}) ${S^\mu}_{; \mu}=(\zeta/T)\theta^2$.
Since $(nU^\mu)_{;\mu}=0$ we have, in the comoving coordinate system, ${S^\mu}_{;\mu}=n\dot{\sigma}$, so that the time derivative of the entropy density becomes
\begin{equation}
n\dot{\sigma}=(\zeta/T)\theta^2.
\end{equation}
\label{13}
The two time derivatives (12) and (13) are seen to be proportional to $\zeta$. Since $\zeta$ is small, we can therefore insert for $a=a(t)$ the expression pertinent for a nonviscous, closed universe: $
a(t)=\sqrt{(8\pi G/3)\rho_0 a_0^4}\,\sin \eta$,
$\eta$ being the conformal time. Imagine now that Eqs.~(12) and (13) are integrated with respect to time. Then, since the densities $\zeta^{-1}\rho a^4$ and $\zeta^{-1} n\sigma$ can be drawn as functions of $t$, it follows that $\rho a^4$ can be considered as a function of $n\sigma$, or, equivalently, that $Ea$ can be considered as a function of $S$. We conclude that this property, previously derived on the basis of CFT, really appears to carry over to the viscous case.

\subsection{Discussion}

We close this section with three remarks:

(i)  The specific entropy $\sigma$ in Eq.~(13) is the usual thermodynamic entropy per particle. The identification of $S$ with $HV/2G$, as made in Eq.~(7), is however something different, since it is derived from  a comparison with the Cardy formula (6).  Since this entropy is the same as the Hubble entropy $S_H$ we can write the equation as 
$ n\sigma_H=H/2G, $
where  $\sigma_H$ is the Hubble entropy per particle. This quantity is different from  $\sigma$, since it does not follow from thermodynamics plus Friedmann equations alone, but from the holographic principle. The situation is actually not peculiar to viscous cosmology. It occurs if $\zeta=0$ also. The latter case is easy to analyze analytically, if we focus attention on the case $t\rightarrow 0$.  Then, for any value of $k$, we have $a \propto t^{1/2}$, implying  that $H=1/2t$. Moreover, from the equation of continuity,  $(nU^\mu)_{;\mu}=0$,  which for a FRW universe yields  $na^3=$ const, so that $n \propto t^{-3/2}$.  The above equation for $\sigma_H$  then yields
    $\sigma_H \propto t^{1/2}$. This is obviously different from the result for the thermodynamic entropy $\sigma$: from Eq.~(13) we simply get $\sigma=$ const when $\zeta=0$. The two specific entropies are thus different even in this case.

(ii)  Our second remark is about the physical meaning of taking the Casimir energy $E_C$ to be {\it positive}. Verlinde assumes that $E_C$ is bounded by the total energy $E$: $
E_C \leq E.$ This may be a realistic bound for some of the CFTs. However, in general cases, 
it is not true.
For a realistic dielectric material with electric dipole interactions it is known that the full Casimir energy is not positive; the dominant terms in $E_C$ are definitely {\it negative}. From a statistical mechanical point of view this follows immediately from the fact that the Casimir force is the integrated effect of the attractive van der Waals force between the molecules. Consider for definiteness the microscopical theory for a dielectric ball. The expression for the Casimir energy contains negative terms as the dominant contributions, but there remains a small residual, cutoff independent, positive term \cite{barton99}. The question emerges: How can such a small residual term in the Casimir energy play a major role in cosmology? Of course our universe is different from a dielectric ball, and we are not claiming that Verlinde's approach is incorrect. Our aim is only to point out that some care should be taken when results from one field of research is applied to another field.

(iii)   Our treatment above was based upon the set of cosmological assumptions $ \{ p=\rho/3,\, k=+1, \,\Lambda=0 \}.$ The recent development of Wang et al. \cite{wang01} is interesting, since it allows for a nonvanishing cosmological constant (still assuming a closed model). One of the scenarios treated in \cite{wang01} is that of a de Sitter universe ($\Lambda >0$) occupied by a universe-sized black hole. A black hole in de Sitter space has the metric
$ ds^2=-f(r)dt^2+f^{-1}(r)dr^2+r^2 d\Omega^2,$
where $f(r)=1-2MG/r-\Lambda r^2/3$. The region of physical interest is that lying between the inner black hole horizon and the outer cosmological horizon, the latter being determined by the magnitude of $\Lambda$. 

Although we do not enter into any detail about this theory, we make the following observations: the above metric is {\it static}; there is no time-dependent scale factor involved, and the influence from viscosity will not turn up in the line element. Moreover, Wang et al. make use of only the member (4) of Friedmann's equations which, as we have noticed, is formally independent of viscosity.

Does this imply that viscosity is without any importance for the present kind of theory? The answer in our opininon is no, since the theory operates implicitly with the concept of the maximum scale factor $a_{max}$ in the closed Friedmann universe. In order to calculate $a_{max}$, one has to solve the Friedmann equation (5) also, which contains the viscosity through the modified pressure $\tilde{p}$. Thus, viscosity comes into play after all, though in an indirect way.

\section{Generalization of Youm's Entropy Formula}

Our argument in section 2.2 was based on the assumption of a radiation dominated universe. Let us now assume that the fluid instead satisfies only the weaker equation of state (1). The paper of Youm \cite{youm02}, mentioned in section 1, is interesting, since it shows that the entropy can no longer be expressed in the usual Cardy-Verlinde form in this case. A more general entropy formula results. And this brings us to the topic of this section, namely to explore to which extent the presence of a bulk viscosity, together with Eq.~(1), influences the entropy formula. In fact, it will turn out that the modified entropy formula found by Youm still persists, even when a constant bulk viscosity is allowed for.

We take the number $n$ of space dimensions to be equal to 3, assume a FRW metric with $k=+1$, and set $\Lambda$ equal to zero. The energy-momentum tensor is given by Eq.~(2), as before (with $\eta =0$), and the Friedmann equations are the appropriate versions of Eqs.~(4) and (5). Equation (12) becomes replaced by
\begin{equation}
\frac{d}{dt}\left( \rho a^{3\gamma}\right)=\zeta\,\theta^2a^{3\gamma},
\end{equation}
\label{14}
while Eq.~(13) remains unchanged.

We can now carry out the same kind of reasoning as above \cite{brevik02a}: 
 Since $\zeta$ is small, we can use for $a=a(t)$ the same expression as given in section 2.2 for a nonviscous closed universe. 
 Imagine that Eqs~(14) and (13) are integrated with respect to time. Since $\zeta^{-1}\rho a^{3\gamma}$ and $\zeta^{-1}n\sigma $ can be drawn as functions of $t$, it follows that $\rho a^{3\gamma}$ can be considered as a function of $n\sigma$. Then, since the total energy is $E \sim \rho a^3$ and the total entropy is $S \sim n\sigma a^3$, it follows that $E a^{3(\gamma-1)}$ is independent of the volume $V$ and is a function of $S$ only. This generalizes the pure entropy dependence of the product $Ea$, found by Verlinde \cite{verlinde00} in the case of a nonviscous radiation dominated universe. And it is noteworthy that the derived property of $Ea^{3(\gamma-1)}$ formally agrees exactly with the property found by Youm \cite{youm02} when $\zeta=0$.

Let us carry out the analysis a bit further, and write the total energy $E$ as a sum of an extensive part $E_E$ and a subextensive part $ E_C$, as we did in section 2.2.
Under a scale transformation $S \rightarrow \lambda S$ and $ V \rightarrow \lambda V $ with constant $\lambda$, $E_E$ scales linearly with $\lambda$. But the term $E_C$ scales with a power of $\lambda$ that is less than one: as $E_C$ is the volume integral over a local energy density expressed in the metric and its derivatives, each of which scales as $\lambda^{-1/3}$, and as the derivatives occur in pairs, the power in $\lambda$ has to be 1-2/3= 1/3. Thus we have
\begin{equation}
E_E(\lambda S, \lambda V)=\lambda E_E(S,V), \quad E_C(\lambda S,\lambda V)=\lambda^{1/3}\,E_C(S,V),
\end{equation}
\label{15}
which implies
\begin{equation}
E_E=(C_1/4\pi) a^{-3(\gamma-1)}\,S^\gamma,\quad E_C=(C_2/2\pi) a^{-3(\gamma-1)}\,S^{\gamma-2/3},
\end{equation}
\label{16}
where $C_1, C_2$ are constants. In CFTs, their product is known: $\sqrt{C_1C_2}=n=3$ \cite{verlinde00}; this being a consequence of the AdS-CFT correspondence. We thus obtain
\begin{equation}
S=\left[ \frac{2\pi a^{3(\gamma-1)}}{\sqrt{C_1C_2}}\sqrt{E_C(2E-E_C)}\right]^{\frac{3}{3\gamma-1}}.
\end{equation}
\label{17}
This is the generalized Cardy-Verlinde formula, in agreement with Eq.~(20) in Youm's paper, reducing to the standard formula (with square root) in the case of a radiation dominated universe. In conclusion, we have extended the basis of Eq.~(17) so as to include the presence of a constant bulk viscosity in the cosmic fluid.

\section*{Acknowledgement}

A substantial part of this work is based on Ref.~\cite{brevik02}, and I thank my co-author, Sergei Odintsov, for his contribution.

\end{document}